\documentclass{acm_proc_article-sp}
\usepackage{graphicx}
\usepackage{fancyvrb}
\usepackage[toc,page]{appendix}

\begin{document}

\title{Concolic Execution as a General Method of Determining Local Malware Signatures}
\subtitle{A preliminary exploration of strategy, feasibility, and potential use cases}

\numberofauthors{1} 
\author{
\alignauthor
Aubrey Alston\\
       \email{ada2145@columbia.edu}
\alignauthor
}

\date{12 December 2015}

\maketitle
\begin{abstract}
A commonly shared component of antivirus suites is a local database of malware
signatures that is used during the static analysis process.  Despite possible
encryption, heuristic obfuscation, or attempts to hide this database from 
malicious end-users (or competitors), a currently avoidable eventuality for offline
static analysis is a need to use the contents of the database in local computation
to detect malicious files.  This work\footnote{This report corresponds to an undergraduate project completed in the Columbia University 
IDS Lab in 2015.} serves as a preliminary exploration 
of the use of concolic execution as a general-case technique for reverse-engineering malware signature database contents: indeed, the existence 
of a practical technique to such an end would certainly require the use 
of true (in the sense of provable security) obfuscation in order for 
malware databases to remain private against capable attackers--a 
major obstacle given the scarcity of truly practical secure obfuscation constructions.  Our work, however, only shows that 
existing tools (at the time of this report) for concolic execution have 
severe limitations which prevent the realization of this strategy.

Given eventual local execution, concolic execution--the (at least partial) treatment
of inputs as symbolic logical variables as they are propagated through a program--
presents itself as a possible means of generating inputs or modifications of inputs
to antivirus software that are capable of triggering (or avoiding) detection by satisfying
the logical constraints required to reach the desired result.

In this project, work has been performed to (1) identify a generally applicable
strategy for applying concolic execution to trigger or avoid static detection in any
antivirus suite, (2) explore the feasibility of this strategy with respect to
current limits to concolic execution, and (3) propose potential schemes to
improve its efficacy.  To achieve these, the devised strategy was applied to
three of the largest concolic execution suites: Angr\cite{angr}, Triton\cite{triton}
, and S2E\cite{s2e} in an attempt to trigger detection in the open-source AV
suite ClamAV in a minimal Ubuntu version 14 environment.

As we have already noted, each framework was noted to have fallen short of 
achieving the desired goal due to both ad-hoc, framework-specific issues 
and larger problems that pervade concolic execution of large systems more generally.  Given the common shortcomings
between the three frameworks, two main blocking areas were identified, and preliminary
suggestions for solutions have been proposed, along with consideration of additional
issues that may arise.
\newline\newline\newline\newline
\end{abstract}

\section*{Background}

\subsection*{Concolic Execution}

\begin{figure}[h!]
  \centering
    \includegraphics[width=0.5\textwidth]{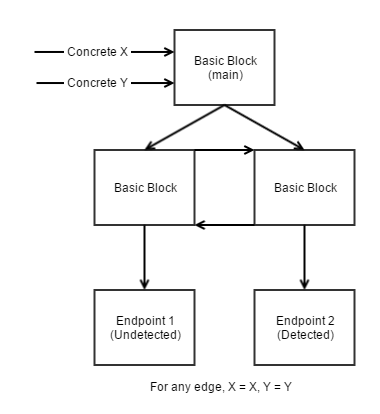}
    \caption{ Conventional Execution }
\end{figure}

Above is a figure of generic conventional execution: a program (say, a
 malware scanner) takes two inputs, X and Y.  Given these inputs, program
 control moves between basic blocks until a program endpoint is reached.

\begin{figure}[h!]
  \centering
    \includegraphics[width=0.5\textwidth]{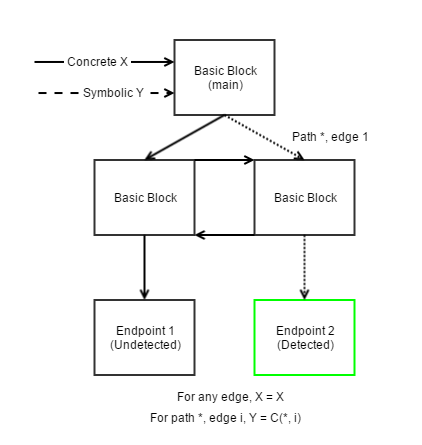}
    \caption{ Concolic Execution }
\end{figure}

As shown above, concolic execution is a modification of conventional
execution in which a subset of inputs are treated as symbolic logical
variables and propagated through the program.  In concolic execution
of a program having more than one possible path, there is no one
deterministic path which the program follows.  Instead, execution
follows a symbolic path, the symbolic variables being augmented according
to operations performed in basic blocks and required state to perform
the necessary branches along the path.

As a hard example, consider the scenario pictured in figure 2.  The
program receives concrete input X and symbolic input Y.  Imagine that there
is a target path of interest \textit{*} from \textit{main} to \textit{Endpoint 2}
as pictured in figure 2.  

At edge 1, \begin{math} Y = C(*,1,Y)\end{math}, which is defined the set of logical
operations performed on Y in main plus the logical constraints on Y required
for control to branch to edge 1.

At edge 2, \begin{math} Y = C(*,2,Y)\end{math}, which is defined as the set of logical
operations performed on C(*,1,Y) in the basic block plus the logical constraints on 
Y required for control to branch to Endpoint 2.

At Endpoint 2, Y is a logical expression, the satisfiable\footnote{ Satisfiability is NP-complete; current symbolic execution engines use approximation to achieve satisfiable assignment. A common technique is to express constraints in the format of satisfiability modulo theorems (SMT) and use Microsoft's Z3 SMT solver.} assignment to which is a program input which would cause the program to reach Endpoint 2 during conventional execution.

\section*{Method}

In order to test the feasibility of using concolic analysis to coerce detection,
the following overarching strategy was applied using three different symbolic
execution frameworks:
\begin{enumerate}
    \item Perform limited manual analysis on the clamscan executable to determine (1) relevant regions of memory to mark as symbolic and (2) the desired (detection) endpoint.
    \item Generate a minimal signature database including only a single hex-match rule for a short string.
    \item Configure and run framework analysis on the clamscan executable using a concrete database file (generated in 2.) and a symbolic file, with the target path endpoint determined in 1.
    \item Collect logical constraint output and generate fake malware (or re-generate the original test string from 2.).
\end{enumerate}

\subsection*{Framework 1: Angr}

Angr is a user-friendly symbolic execution framework which abstracts the entire process of symbolic execution.  It achieves symbolic execution via the following workflow:
\begin{enumerate}
    \item Load binary to be analyzed.
    \item Translate the binary to an intermediate representation.
    \item Translate the intermediate representation to a semantic representation.
    \item Perform analysis.
\end{enumerate}

In order to perform instrumentation and perform binary analysis, a user must
write scripts using Angr's Python bindings on a one-off basis.

The following is the following pattern that an Angr script follows:

\verb|# Load binary|\newline
\verb|b_file = angr.Project(file_location)|\newline\newline
\verb|# Specify starting instruction|\newline
\verb|start_addr = <addr>|\newline
\verb|state = binary_file.factory.blank_state(start_addr)|\newline\newline
\verb|# Hook troublesome instructions to predefined replacements|\newline
\verb|binary_file.hook(<addr>, <function>, length=instr_length)|\newline\newline
\verb|# Mark memory as symbolic|\newline
\verb|state.memory.store(<addr>, state.se.BV(<name>, <bit len>))|\newline\newline
\verb|# Perform analysis, seeking a path containing an instruction|\newline
\verb|p_group = b_file.factory.path_group(state)|\newline
\verb|p_group.explore(find=<addr>)|\newline

\subsubsection*{Results and Issues}

Angr proved incapable in its current form to instrument and analyze the ClamAV
clamscan binary.  Because Angr lifts everything to an intermediate representation
at the application level, Angr struggles to simulate marking OS-level
components (such as files) as symbolic; additionally, Angr is not able to handle
symbolic execution of dynamic library calls in its current state and would find 
difficulty in doing so because it attempts to analyze the entire binary at once.

The major limit of Angr is that its transparency is limited purely to the binary in question.  This generally shouldn't be a problem for small binaries, but for a system as complicated as a general-case antivirus suite, a suitable symbolic execution framework must offer additional symbolic reach.

\subsection*{Framework 2: Triton}

Triton is a symbolic execution framework which achieves symbolic execution by performing symbolic execution and taint analysis on top of conventional execution by instrumenting the binary with Pintool.  Because Triton uses Pintool, it is more transparent than Angr in that it instruments dynamically and has the potential (due to the use of Pintool) to eventually interface with dynamically generated/loaded code and in that it actually natively executes the binary.

Triton achieves symbolic execution via the following workflow:
\begin{enumerate}
    \item Invoke the binary using Pin and a special Triton Pintool which handles symbolic execution.
    \item Internally update constraints and perform symbolic operations ad-hoc during the instrumented execution of the target binary.
\end{enumerate}

Like Angr, a user must write scripts using Triton's Python bindings on a one-off basis, invoking the script through Triton's packaged, modified version of Pin.

In order to perform instrumentation and perform binary analysis, a user must
write scripts using Angr's Python bindings on a one-off basis.

The following is the following pattern that a Triton script follows:

\verb|# Only perform actions when invoked as main|\newline
\verb|if __name__ == '__main__':|\newline
\verb|    # Specify address from which analysis should start|\newline
\verb|    startAnalysisFromAddr(<addr>)|\newline
\verb|    # Add callbacks(this is where you check addresses, symbolic memory, etc |\newline
\verb|    addCallback(<function>, IDREF.CALLBACK.AFTER)|\newline
\verb|    # Start analysis|\newline
\verb|    runProgram()|\newline

Once written, the Triton Python script is invoked as follows:

\verb|sudo ./triton clamav_extraction.py clamscan|\newline
\verb|       -d testdb.db <file>|\newline

\subsubsection*{Results and Issues}

Triton also proved incapable in its current form to instrument and analyze the ClamAV
clamscan library.  Triton currently does not provide for use of Pintool's support
for instrumentation of dynamically loaded code; additionally, despite offering slightly more transparency than Angr, Triton still doesn't provide a coherent means of marking OS-level objects (such as files) as symbolic.  

Triton's greatest weakness is its incompleteness in implementation.  It is currently not using Pintool to its fullest potential, and it additionally faces trivial (yet critical) bugs, such as out-of-memory errors and segmentation faults when attempting to access a symbolic region of memory\footnote{Triton managed to bleed through Oracle VMWare's Hypervisor and crash my laptop when I tried to access a single byte of symbolic memory.}

\subsection*{Framework 3: S2E}

S2E (Selective Symbolic Execution) is a heavyweight symbolic execution framework
that offers a lot of power with respect to dynamically instrumenting binaries.

S2E goes one step deeper than Triton, achieving symbolic execution as follows:

\begin{enumerate}
    \item Configure plugins for achieved functionality and output for symbolic execution on the machine.
    \item Initialize and run a virtual machine--using a modified implementation of QEMU-- that introduces custom opcodes to perform symbolic execution at the hardware level.
    \item Once started, use S2E utilities to pull instrumentation utilities from the host VM to the guest VM.
    \item Use the main instrumentation utility, \verb|init_env.so|, to invoke the target binary as a blackbox with the desired concrete and symbolic arguments from the guest VM, performing the symbolic execution on the host with KLEE.
\end{enumerate}

S2E diverges from Triton and Angr in that it does not encourage or
require writing scripts on a one-off basis: while it is possible to write new plugins for S2E, their functionality is expected to be generally applicable to all instrumented binaries--not just one.  This gives it a significant advantage
over the other two frameworks for the purpose of this project in that a solution
for one antivirus software suite would hypothetically work for virtually all of them.

Another significant advantage to the S2E framework is that it does allow for symbolic representation of OS-level objects, providing utilities out of the box for writing and working with symbolic files.  With respect to the promises made by the framework, S2E is the most promising framework to house this project.

In order to use S2E to coerce detection using symbolic execution, the following workflow should be followed:

\begin{enumerate}
    \item Attain and put minimal S2E configuration in place.  Ensure that the TestCaseGenerator plugin is present (this is a provided plugin which allows for symbolic constraints to be directly converted to concrete examples as soon as divergent paths are found).
    \item Start QEMU in S2E mode.
    \item Using the S2E s2eget utility, pull instrumentation utiltiies from the host VM.
    \item Use the S2E s2ecmd utility to write a symbolic file.
    \item Invoke clamscan using \verb|init_env.so| with the concrete minimal malware database file and the symbolic file created in the previous step.
    \item Collect and analyze results from the host VM.
\end{enumerate}

\subsubsection*{Results and Issues}

For all the promising aspects of S2E, it still has some shortcomings which prevent complete application of the strategy described for coercing detection in ClamAV.  S2E is robust enough to handle the complete symbolic execution of clamscan; however, S2E, despite having the capacity, does not currently implement any way to handle symbolic execution including dynamic library calls.

S2E also falls short in terms of documentation and disclosure with respect to what features are completely implemented.  For example, S2E will follow through and attempt to execute clamscan symbolically; however, it handles its inability
to execute dynamic library calls symbolically by silently concretizing symbolic
arguments to the first possible example, disallowing any real results to be obtained.  Additionally, S2E natively provides no way to select desired target branches (something Triton and Angr do offer).

In spite of these issues, S2E provides the space and tools to address these issues with more ease than Angr or Triton.  If a solution is devised for the dynamic library issue, it could easily be implemented as a modification to\verb|init_env.so|; to handle S2E's issues with path selection, one could (given the time) implement a plugin to cover this shortcoming.

\section*{Discussion and Conclusions}

Of all three frameworks, S2E presents itself as being best-suited to the task.  Nevertheless, there are still two main problem areas to be addressed before the method of this project can be fully realized.

\subsection*{Issue: Dynamic Libraries}

An issue faced by all three frameworks was the issue of dynamic libraries.  None of these frameworks in their current state are capable of handling dynamic libraries.  In the case of Angr and Triton, execution simply halts and crashes; however, in the case of S2E, the situation is somewhat misleading.

\begin{figure}[h!]
  \centering
    \includegraphics[width=0.5\textwidth]{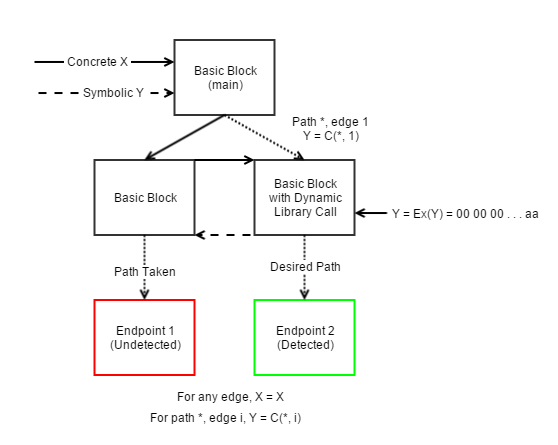}
    \caption{ Handling of External Library Call in S2E }
\end{figure}

As shown in Figure 3, S2E silently concretizes symbolic inputs to external library
functions.  On return, the symbolic variable is now the concretized value
from before the system call, preventing control from reaching the 
desired path endpoint.

\subsection*{Issue: Path Explosion and Path Selection Heuristics}

An overarching problem facing symbolic and concolic execution in general is the path selection problem.  To illustrate what this is, exactly, consider the basic block structure given in figures 1, 2, and 3.  If the symbolic execution engine attempts to explore all feasible paths of the program, and the loop between the two central basic blocks is caused by a \verb|while(true)|, then there could be an infinite number of paths leading from \verb|main()| to either endpoint, and the engine might never feasibly reach one endpoint over the other.

Concolic execution mitigates this issue to an extent by concretizing as much input space as possible; however, with the number of possible endpoints and paths within a system as large as an antivirus suite, heuristics must be available to select one endpoint over another.  

S2E does not provide such a means of path selection out of the box, but it does provide the components necessary to be able to write plugins to achieve the desired functionality (Angr and Triton both implement some sort of heuristic to achieve these, so an initial solution may be a port of these).  As it stands, when analysis is attempted in S2E, between this and the previous issue, S2E simply is not able to reach the desired program endpoint.

\subsection*{Future direction}

In order to further the goals of this project, the above issues must be resolved.  S2E provides the technical tools and capability required; however, plugins and modifications will likely need to be implemented to fill the gaps.

A possible solution to both of these issues may lie in the replay functionality KLEE implements (but S2E currently doesn't use).

\begin{figure}[h!]
  \centering
    \includegraphics[width=0.5\textwidth]{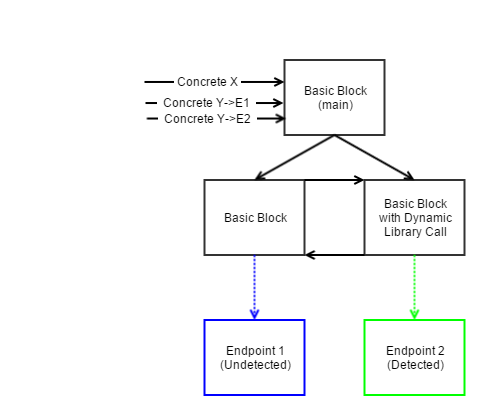}
    \caption{ Solution Sketch: Pre-Analysis Map Execution }
\end{figure}

Figure 4 shows a possible additional step before the analysis issue which may contribute to resolving both issues currently preventing progress towards this project's goal.  Before the true symbolic execution of the binary of interest, provide it first with a set of concrete inputs known to reach points of execution of interest.  For each of these inputs, execute the program once with the input, and then use KLEE's replay functionality to map (1) where dynamic code is loaded and (2) the basic blocks reached during execution using the inputs of interest.  Then, while performing symbolic execution, instrument dynamic calls to reuse the dynamically loaded code from the external libraries (without concretizing symbolic inputs), limiting path exploration to those including points reached during the execution of the pre-run inputs.

It is suggested that this project be continued by implementing new plugins for S2E and possibly by modifying \verb|init_env.so| to achieve the above.  With these additional components, S2E could be used to further explore the feasibility and effectiveness of determining information about malware signatures in antivirus software using concolic execution. 

\bibliographystyle{abbrv}


\end{document}